\documentclass[prl,aps,twocolumn,showpacs]{revtex4-1}
\usepackage{amsmath,amssymb,graphicx}
\usepackage{epsfig}
\usepackage{textcomp}
\usepackage{color}
\usepackage{cases}
\usepackage{epstopdf}
\usepackage{multirow}
\usepackage{array}
\usepackage{hyperref}
\hypersetup{pdfpagemode=UseNone,colorlinks=true,linktoc=page,pdfborder={0 0 0},linkcolor=blue,citecolor=blue}

\begin{document}
\title{Effect of reaction step-size noise on the switching dynamics of stochastic populations}

\author{Shay Be'er, Metar Heller-Algazi, and Michael Assaf}

\affiliation{Racah Institute of Physics, Hebrew University
of Jerusalem, Jerusalem 91904, Israel}


\begin{abstract}
In genetic circuits, when the mRNA lifetime is short compared to the cell cycle, proteins are produced in geometrically-distributed bursts, which greatly affects the cellular switching dynamics between different metastable phenotypic states. Motivated by this scenario, we study a general problem of switching or escape in stochastic populations, where influx of particles occurs in groups or bursts, sampled from an arbitrary distribution. The fact that the step size of the influx reaction is a-priori unknown, and in general, may fluctuate in time with a given correlation time and statistics, introduces an additional \textit{non}-demographic \textit{step-size} noise into the system. Employing the probability generating function technique in conjunction with Hamiltonian formulation, we are able to map the problem in the leading order onto solving a stationary Hamilton-Jacobi equation. We show that bursty influx exponentially decreases the mean escape time compared to the ``usual case" of single-step influx. In particular, close to bifurcation we find a simple analytical expression for the mean escape time, which solely depends on the mean and variance of the burst-size distribution. Our results are demonstrated on several realistic distributions and compare well with numerical Monte-Carlo simulations.
\end{abstract}

\maketitle


Stochastic populations possessing multiple long-lived metastable states can undergo noise-induced switching, which plays a key role in many systems in physics, chemistry, biology, and ecology~\cite{Horsthemke,Hanggi,Gardiner,Balazsi}. Many previous studies of such noise-induced switching or escape have focused on the role of demographic noise, see \textit{e.g.}, Refs.~\cite{Bartlett,Nisbet,Dykman,KS,explosion,EK,Dykman1,PRE2010,OMII,ours}. Here a stochastic population, which regulates itself through random births and deaths, eventually escapes its initial metastable state via a rare sequence of events caused by the demographic noise.

In models of stochastic population dynamics, an important process that prevents the population from going extinct and is the main driver towards establishment of the population is the process of immigration or influx of particles~\cite{ours,Brown,Hanski}. In most cases, immigration is modeled using single-step influx (SSI). Here, influx occurs via a single-step process  $\emptyset\to A$, which corresponds to a single individual arriving at a certain rate, increasing the population size from $n$ to $n+1$ individuals.

Yet, a more general and realistic way to model immigration is to consider bursty influx (BI) of particles or individuals. Here, influx occurs in bursts of individuals, with the burst size sampled from an arbitrary burst-size distribution (BSD). This generalized BI process appears in various scientific areas. In ecology, BI can account for more complex migration patterns such as arrival in groups~\cite{Goel,OMII}. In gene expression, when the mRNA lifetime is short compared to the cell cycle, protein synthesis effectively occurs in geometrically-distributed bursts~\cite{Arkin,PaulssonI,Paulsson,RajI,Suter,Swain,XieI,XieII,AssafRobertsSchulten,Earnest}. This gives rise to non-Poissonian protein statistics~\cite{Arkin,Paulsson,RajI,Swain} and may exponentially decrease switching times between different phenotypic states~\cite{Arkin,Suter,MickeyElijah,AssafRobertsSchulten,Earnest}. Moreover, in evolutionary biology it has been shown that rapid punctuational bursts of speciation is a significant mechanism responsible for the evolution of species~\cite{Pagel}. A similar mechanism of BI of lexical changes is also key in evolutionary linguistics~\cite{Pagel1}.

Importantly, accounting for uncertainty in the reaction step size introduces additional \textit{non}-demographic noise into the system, which can be called  \textit{step-size} noise, and has not been systematically investigated before. This noise, in general, includes a step-size parameter, $k(t)$, fluctuating in time, with a given statistics and correlation time. Incorporating reaction step-size noise is expected to dramatically change the statistics of interest. This is since, \textit{e.g.}, rare events such as switching or escape, are expected to be determined by a competition between the tails of the probability distribution of the population sizes given a certain burst size, and the tails of the BSD.

In this paper, we make a first step towards a systematic investigation of generic reaction step-size noise. We do so by considering static step-size noise in the form of BI, $\emptyset\to kA$, where $k$ can take random integer values drawn from an arbitrary BSD. We demonstrate our formalism on a specific Verhulst-like model which includes, apart from BI, binary reproduction and death. This model gives rise to two fixed points at the deterministic level: a lower stable point $n_1$, and a higher unstable point $n_2$, see below. As a result, the stochastic population which initially resides in the vicinity of the long-lived metastable state $n_1$, ultimately crosses the unstable point $n_2$ and escapes to infinity, leading to a Malthusian catastrophe~\cite{Bartlett}. In order to study how BI affects the mean escape time (MET) as function of the BSD, we employ the probability generating function formalism~\cite{Gardiner}. By doing so, we transform the master equation into a single partial differential equation. To this end, we use the WKB theory to treat this equation and find the MET in the leading exponential order as function of the BSD. Our analysis shows that BI can exponentially decrease the MET compared to the usual SSI case. In the bifurcation limit where the stable and unstable fixed points are close, we derive a simple expression for the reduction of the MET, which depends solely on the first and second moments of the BSD. Our results are demonstrated on several BSD examples such as the geometric, negative-binomial, and scale-free distributions, all of which compare well with numerical Monte-Carlo simulations.


Let $n(t)$ be the population size. Our model includes binary reproduction $2A\xrightarrow{\alpha/N}3A$ and death $A\xrightarrow{1}\emptyset$, where $\alpha\gtrsim1$ is the reproduction rate, $N\gg1$ is related to the critical population size, see below, and time is rescaled by the death rate. Accounting for immigration via a single-step influx (SSI) process $\emptyset\xrightarrow{k_0} A$, where $k_0$ is the immigration rate, we obtain the following deterministic rate equation for the mean population size $\bar{n}(t)$
 \begin{equation}
	\label{a}
	\dot{\bar{n}}=k_0+\frac{\alpha\bar{n}^2}{N}-\bar{n}.
\end{equation}

In order to study bursty influx (BI) we replace the SSI process by an infinite set of influx processes $\emptyset \to kA$ ($k=0,1,2,....$), where the burst size, $k$, is drawn from an arbitrary BSD, $D(k)$, which satisfies $\sum_{k=0}^{\infty}D(k)=1$, and has a finite mean $\langle k\rangle$ and variance $var(k)$. The rate of the BI processes $\emptyset \to kA$ is taken to be $k_0D(k)/\langle k\rangle$. This insures, see below, that the rate equation~(\ref{a}) and the corresponding fixed points are unaffected by the BI.

Eq.~(\ref{a}) has two fixed points. The lower, $n_1/N=(1-\sqrt{1-4\alpha k_0/N})/(2\alpha)$, is an attracting fixed point, and the higher, $n_2/N=(1+\sqrt{1-4\alpha k_0/N})/(2\alpha)$, is a repelling fixed point, where $n_2>n_1$, and we assume that $n_1\gg 1$. This system describes a runaway model. Here an initial population, $n_0<n_2$, first relaxes into the long-lived metastable state, $n_1$. Yet, after an exponentially long waiting time, the population crosses the barrier at the critical population size $n_2$, and reaches a state of unlimited growth~\cite{cubic}. Our aim here is to investigate how the MET depends on the functional form of $D(k)$.

To do so, we account for demographic stochasticity and BI, and consider the corresponding master equation for $P_n(t)$ -- the probability to find $n$ individuals at time $t$
\begin{align}
	\label{b}
	\dot{P}_n  = \ & \frac{k_0}{\langle k\rangle}\left[\sum_{k=0}^nD(k)P_{n-k}-\sum_{k=0}^{\infty}D(k)P_n\right]  \\ &  + \frac{\alpha (n-1)^2}{N}P_{n-1}-\frac{\alpha n^2}{N}P_n+(n+1)P_{n+1}-nP_n, \nonumber
\end{align}
where the terms in square brackets correspond to the BI. Multiplying Eq.~(\ref{b}) by $n$ and summing over all values of $n$, we recover rate equation~(\ref{a}), justifying a-posteriori the rate chosen for the BI process. That Eq.~(\ref{b}) includes reactions with all possible step sizes renders this problem difficult to treat via the real-space approach~\cite{Dykman,KS,explosion,EK,Dykman1,PRE2010}. We thus use the momentum-space approach~\cite{Elgart,AssafMeerson,Hydrogen} instead.

We now define the probability generating function of the population, $G=\sum_{n=0}^{\infty}p^nP_n$~\cite{Gardiner}, where $p$ is an auxiliary variable which plays the role of the momentum~\cite{Elgart}. Multiplying Eq.~(\ref{b}) by $p^n$, and summing over all values of $n$ we arrive at a single evolution equation for $G(p,t)$~\cite{FnII}
\begin{align}
	\label{c}
	\partial_tG  = \ & \frac{k_0}{\langle k\rangle}[g(p)-1]G+\left(\frac{\alpha}{N}p-1\right)(p-1)\partial_pG \nonumber \\ &  +\frac{\alpha}{N}p^2(p-1)\partial_p^2G.
\end{align}
Here we have introduced $g(p)=\sum_{k=0}^{\infty}p^kD(k)$. This function serves as the burst-size probability generating function and encodes the arbitrary (known) BSD.

As the MET is exponentially large, see below, once the system has settled into the vicinity of the long-lived metastable state $n_1$ we can replace $\partial_tG$ by zero in Eq.~(\ref{c}). The quasi-stationary solution of Eq.~(\ref{c}), $G_{qs}(p)$, with proper boundary conditions encodes the quasi-stationary distribution of the population sizes around $n_1$, which yields the MET in the leading order~\cite{Elgart,AssafMeerson,Hydrogen}, whereas subleading order calculations require spectral analysis of Eq.~(\ref{c})~\cite{AssafMeerson}. The ordinary differential equation for $G_{qs}(p)$ is, in general, unsolvable analytically, and we thus use the WKB theory in the spirit of~\cite{Hydrogen}. Employing the WKB ansatz $G_{qs}(p)=e^{-NS(p)}$ in Eq.~(\ref{c}), where $S(p)$ is the action, we obtain in the leading $\mathcal{O}(N)$ order, a stationary Hamilton-Jacobi equation, $\mathcal{H}[p,-S'(p)]=0$, with
\begin{equation}
	\label{d}
	\mathcal{H}(p,q)=(p-1)\left[\kappa_0 f(p)-q+\alpha p^2q^2\right].
\end{equation}
Here we have defined $\kappa_0=k_0/N$ and $f(p)\equiv[g(p)-1]/[\langle k\rangle(p-1)]$. Note, that $f(p)$ does not diverge at $p=1$ and satisfies $f(1)=1$. In~(\ref{d}) we have further introduced $q(p)=-S'(p)$, which plays the role of the reaction coordinate conjugate to the momentum $p$~\cite{Elgart,Hydrogen}.

\begin{figure}[b]
\centering
\includegraphics[trim = .35in 1.25in .4in .65in,clip,width=3.25in]{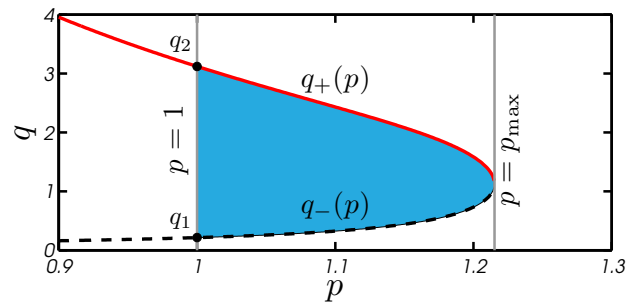}
\caption{(Color online) Shown are characteristic activation trajectories $q_{+}(p)$ (solid line) and $q_{-}(p)$ (dashed line) given by Eq.~(\ref{e}), in the case of geometric BSD. The shaded area represents the accumulated action, see Eq.~(\ref{f}). Here parameters are $\kappa_0=0.2$, $\alpha=0.3$, and $b=3$, see Table~\ref{tab-a}.}
\label{fig-a}
\end{figure}

The equation $\mathcal{H}=0$ has three solutions; one of them, $p=1$, is the trivial solution called the relaxation trajectory, and corresponds to the deterministic dynamics~\cite{Elgart}. Indeed, at $p=1$, the Hamilton equation for $\dot{q}$ satisfies $\dot{q}=\kappa_0+\alpha q^2-q$, which coincides with rate equation~(\ref{a}), upon defining $q=n/N$ as the population concentration. The Hamiltonian, Eq.~(\ref{d}), has two more, non-trivial, zero-energy trajectories, called the activation trajectories
\begin{equation}
	\label{e}
	q_{\pm}(p)=\frac{1\pm\sqrt{1-4\alpha\kappa_0p^2f(p)}}{2\alpha p^2}.
\end{equation}

Figure~\ref{fig-a} shows characteristic activation trajectories between the two fixed points, $q_{1,2}=n_{1,2}/N$, in the case of geometric BSD. Having found the activation trajectories, the MET $\tau$ is given in the leading order by~\cite{Elgart,accuracy}
\begin{equation}
	\label{f}
	\tau\sim e^{N\Delta S}\;,\;\;\;\;\Delta S=\int_1^{p_{\mathrm{max}}}[q_{+}(p)-q_{-}(p)]dp.
\end{equation}
The accumulated action, $\Delta S$, corresponds to the area between the activation and relaxation trajectories, see Fig.~\ref{fig-a}, and $q_{+}(p)$ and $q_{-}(p)$ are determined by Eq.~(\ref{e}). In addition, $p_{\mathrm{max}}$ is determined by the condition $q_{+}(p_{\mathrm{max}})=q_{-}(p_{\mathrm{max}})$ which yields a transcendental equation  $p^2_{\mathrm{max}}f(p_{\mathrm{max}})=1/(4\alpha\kappa_0)$. The condition for $p_{\mathrm{max}}$ stems from the monotonicity in $p$ of the trajectories. Eq.~(\ref{f}) for the MET together with Eq.~(\ref{e}) in the case of an arbitrary BSD is one of our main results.

\begin {table}[b]
\footnotesize
\begin{center}
\begin{tabular}{ | p{.5cm} || p{3.3cm} | p{1cm} | p{1.2cm} | p{1.8cm} |}
	\hline
& \multirow{2}{*}{$D(k)$} & \multirow{2}{*}{$\langle k\rangle$} & \multirow{2}{*}{$\mathrm{var}(k)$} & \multirow{2}{*}{$f(p)$} \\ [0.25cm]
	\hline
\multirow{2}{*}{SSI} & \multirow{2}{*}{$\delta_{k,1}$} & \multirow{2}{*}{$1$} & \multirow{2}{*}{0} & \multirow{2}{*}{$1$} \\ [0.2cm]
	\hline
\multirow{2}{*}{KSI} & \multirow{2}{*}{$\delta_{k,K}$} & \multirow{2}{*}{$K$} & \multirow{2}{*}{0} & \multirow{2}{*}{$\frac{p^K-1}{K(p-1)}$} \\ [0.3cm]
	\hline
\multirow{2}{*}{GE} & \multirow{2}{*}{$\left(\frac{b}{1+b}\right)^k\left(\frac{1}{1+b}\right)$} & \multirow{2}{*}{$b$} & \multirow{2}{*}{$b(b+1)$} & \multirow{2}{*}{$\frac{1}{1+b-bp}$} \\ [0.3cm]
	\hline
\multirow{2}{*}{NB} & \multirow{2}{*}{${k+a-1\choose k}\left(\frac{1}{1+b}\right)^a\left(\frac{b}{1+b}\right)^k$} & \multirow{2}{*}{$ba$} & \multirow{2}{*}{$b(b+1)a$} & \multirow{2}{*}{$\frac{(1+b-bp)^{-a}-1}{ba(p-1)}$} \\ [0.3cm]
	\hline
\multirow{2}{*}{SF} & \multirow{2}{*}{$k^{-\gamma}/H_N^{(\gamma)}, \ 1< k\leq N$} & \multirow{3}{*}{$\frac{H_N^{(\gamma-1)}}{H_N^{(\gamma)}}$} & \multicolumn{2}{l |}{\multirow{3}{*}{$\frac{H_N^{(\gamma-2)}}{H_N^{(\gamma)}}-\langle k\rangle^2$ $\Bigg|$ see~\cite{fequation}}}  \\ [0.3cm]
\multirow{2}{*}{} & \multirow{1}{*}{$0, \ \ \ \ \ \ \ \ \ \ \ \ \ k>N$} & \multirow{2}{*}{} & \multicolumn{2}{l |}{} \\ [0.05cm]
 \hline
\end{tabular}
\caption{The burst-size distribution (BSD), its mean, variance, and the corresponding $f(p)$ [defined immediately after Eq.~(\ref{d})] for single step influx (SSI), $K$-step influx (KSI), geometric (GE), negative-binomial (NB), and scale-free (SF) BSDs. Here, $H_N^{(\gamma)}$ is the $N$-th harmonic number of order $\gamma$.}
\label{tab-a}
\end{center}
\end {table}


To evaluate the effect BI has on the MET we compare Eq.~(\ref{f}) to the simple case of SSI, see Table~\ref{tab-a}. Here, $\Delta S$ from Eq.~(\ref{f}) can be explicitly computed and we find
\begin{equation}\label{ssi}
\hspace{-2.5mm}\Delta S_{SSI}=\sqrt{\frac{\kappa_0}{\alpha}} \left[\sqrt{\frac{1}{\alpha\kappa_0}-4}+2\arcsin(2\sqrt{\alpha\kappa_0})-\pi\right].
\end{equation}
Therefore, the general result for $\tau$ for an arbitrary BSD can be written as $\tau\sim e^{N\Delta S}=(\tau_{SSI})^{\phi}$, where we define $\phi\equiv\Delta S/\Delta S_{SSI}$ as the ratio between the accumulated actions in the BI and SSI cases. In the following we calculate the exponent $\phi$ for various BSDs.

\begin{figure}[t]
\centering
\includegraphics[trim = .15in .8in .5in .85in,clip,width=3.4in]{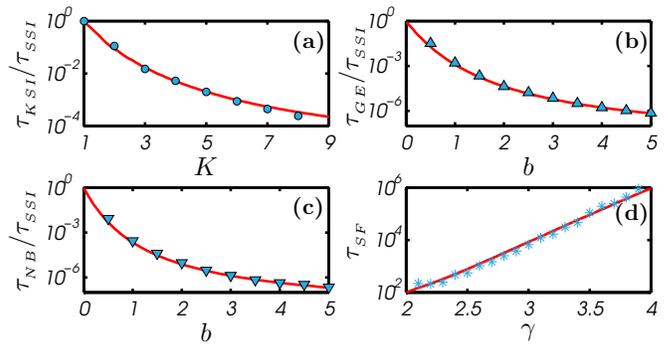}
\caption{(Color online) Panels (a-c): ratio between the MET for various BSDs and the MET for the SSI case as a function of different characteristic parameters, see Table~\ref{tab-a}. Solid line - theoretical result ~(\ref{f}); markers - simulation results. Panel (d): the MET for SF distribution as a function of $\gamma$. In all panels, we have multiplied our analytical results by constant pre-exponential factors obtained by a numerical fit: $21$, $14$, $10$, $9$, $2$ for single-step influx, $K$-step influx, geometric, negative-binomial, and scale-free BSDs, respectively. In all panels $k_0=10$ and $\alpha=1.1$; panel (a): $N=90$; panel (b): $N=105$; panel (c): $N=105$ and  $a=2$; panel (d): $N=250$.}
\label{fig-b}
\end{figure}

Figure~\ref{fig-b} compares between the theoretical and numerical results for the MET, for different BSDs defined in Table~\ref{tab-a}. Here, our numerical simulations were performed using an extended version of the Gillespie algorithm~\cite{Gillespie} that accounts for BI. For all BSDs considered, an exponential reduction of the MET is observed. To remind the reader, we have chosen the BI rate in our model in such a way that the deterministic description remains unchanged compared to the SSI case. Therefore, this reduction in the MET is a net effect of BI and occurs due to the right tail of the BSD. The latter encodes occasional large bursts of individuals, which allow for a more rapid escape from the metastable state.

Let us now discuss in detail the different BSDs we have used. We begin with the simple case of influx occurring in bursts of constant size $K>1$, see Table~\ref{tab-a}. While this BSD differs from the SSI case only by the mean, the fact that influx occurs in bursts of size $K>1$ at a rate $k_0/K$ suffices to reduce the MET exponentially, see  Fig.~\ref{fig-b}a. As expected, at $K=1$ the $K$-step case reduces to $\tau_{SSI}$.

Next, we consider the case of geometric BSD, which occurs, \textit{e.g.}, in cell biology \cite{XieI,XieII,Raj,Swain,Suter}, and also the more general case of negative-binomial BSD, whose mean and variance can be independently controlled. In both these cases, one observes an exponential reduction in the MET compared with the SSI case, see Fig.~\ref{fig-b}b+c. As expected, see Fig.~\ref{fig-b}b+c, for $b\to 0$ for which the width of the geometric and negative binomial BSDs vanishes, see Table~\ref{tab-a}, the  MET reduces to the SSI case, since $f(p)\to 1$.

Until now we have considered BSDs with an ${\cal O}(1)$ variance. To fully explore the effect of BI we have also studied the case of bursts sampled from a scale-free distribution, $D(k)\sim k^{-\gamma}$, which plays a key role in many processes in nature~\cite{BarabasiAlbert}. Here, for $\gamma\leq 3$, the BSD variance scales with the system size. Thus, we apply a natural cutoff for the scale-free BSD, see Table~\ref{tab-a} and~\cite{cutoff}.
%
In this case we also observe excellent agreement between the analytical and numerical results, see Fig.~\ref{fig-b}d~\cite{fequation,FnIX}.


In the bifurcation limit the results drastically simplify. Here one can find an analytical expression for the ratio of the accumulated actions, $\phi$, as a function of the first two moments of the BSD. Let us define the bifurcation parameter $\varepsilon\equiv \sqrt{1-4\alpha\kappa_0}\ll 1$, which guarantees that the fixed points are close $q_2-q_1\sim\varepsilon$. This limit allows for the parabolic approximation of the activation trajectories for which $p-1\sim\varepsilon^2$~\cite{FnVIII}. Expanding $g(p)$ around $p=1$, we find $f(p) = (1/\langle k\rangle)\sum_{m=1}^{\infty}g^{(m)}(1)(p-1)^{m-1}/m!$, where $g^{(m)}(1)$ is the $m$-th derivative of $g(p)$ evaluated at $p=1$, which yields combinations of BSD moments. Retaining the leading- and subleading-order terms yields
\begin{equation}
	\label{i}
	f(p) \simeq 1+2\Omega(p-1)\;;\;\;   \Omega=\frac{1}{4}\left[\mathrm{var}(k)/\langle k\rangle+\langle k\rangle-1\right],
\end{equation}
where $\langle k \rangle$ and $\mathrm{var}(k)$ are the mean and variance of the BSD. Here, the leading order term of $f(p)$ corresponds to the SSI result, see Table~\ref{tab-a}, while the subleading order term encodes the effect of BI for any arbitrary BSD.

Defining a rescaled momentum coordinate $\tilde{p}=(p-1)/\varepsilon^2$, $\Delta S$ [Eq.~(\ref{f})] becomes
$\Delta S=\varepsilon^2\int_0^{\tilde{p}_{\mathrm{max}}}[q_{+}(\tilde{p})-q_{-}(\tilde{p})]d\tilde{p}$. Using Eq.~(\ref{i}), in the leading order in $\varepsilon\ll 1$, the integrand becomes $q_{+}(\tilde{p})-q_{-}(\tilde{p})\simeq 4\kappa_0\varepsilon\sqrt{1-2\tilde{p}(1+\Omega)}$, while $\tilde{p}_{\mathrm{max}}\simeq 1/[2(1+\Omega)]$. Plugging these results into the integral for $\Delta S$ and preforming the integration, we find $\Delta S$ in the leading order in the bifurcation limit $\varepsilon\ll 1$: $\Delta S^b\simeq\Delta S^b_{SSI}/(1+\Omega)$. Here, using Eq.~(\ref{ssi}), $\Delta S^b_{SSI}\simeq 4\kappa_0\varepsilon^3/3$ is the result for the SSI case in the bifurcation limit. As a result, close to bifurcation, the ratio of accumulated actions reads
\begin{equation}
	\label{k}
	\phi^b=\Delta S^b/\Delta S^b_{SSI}=(1+\Omega)^{-1}.
\end{equation}
The validity of the results for $\Delta S^b$ and $\Delta S^b_{SSI}$ can be checked a-posteriori. Indeed, the WKB theory formally requires that $N\Delta S\gg1$~\cite{Dykman}. That is, $\varepsilon$ has to satisfy $N^{-1/3}\ll\varepsilon\ll 1$ for the analytical results to be accurate. Also, note that due to the Cauchy-Schwarz inequality, $\Omega$, defined by Eq.~(\ref{i}) satisfies $\Omega\geq 0$. Only in the case of SSI, for which $\langle k \rangle=1$ and $\mathrm{var}(k)=0$, we have $\Omega=0$ and $\phi^b=1$. Thus, close to bifurcation, we have analytically shown that BI of any arbitrary BSD exponentially reduces the MET. We have numerically confirmed that this result also holds beyond the bifurcation limit.

\begin{figure}[t]
\centering
\includegraphics[trim = .25in .45in .4in .9in,clip,width=3.5in]{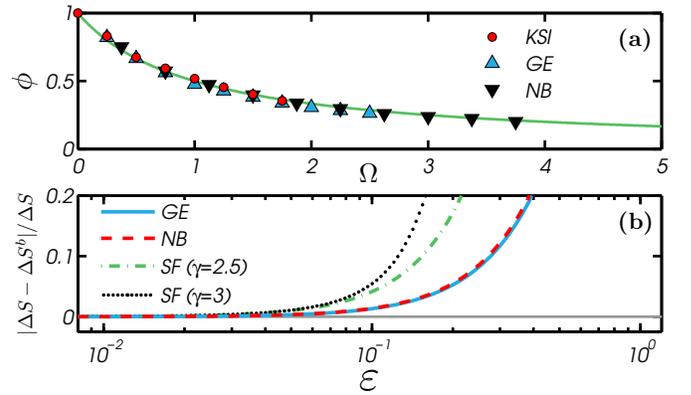}
\caption{(Color online) Panel (a): The accumulated action ratio $\phi$, obtained from numerical simulations (markers) compared with the theoretical bifurcation result, $\phi^b$ [Eq.~(\ref{k})] (solid line). Parameters and data are the same as in Fig.~\ref{fig-b}a-c, and $\varepsilon\simeq 0.75$. Panel (b): Relative error of the accumulated action in the bifurcation limit versus the general case, $|\Delta S-\Delta S^b|/\Delta S$, as a function of $\varepsilon$. Here $\kappa_0=0.2$ for all lines; GE: $b=3$; NB: $b=3$, $a=5$; SF: $N=1000$.}
\label{fig-c}
\end{figure}

Using the numerical simulation results from Fig.~\ref{fig-b}a-c, we compare in Fig.~\ref{fig-c}a the numerical value of $\phi$ with the theoretical bifurcation-limit result, $\phi^b$~(\ref{k}). The numerical results for three different BSDs exhibit a collapse onto the theoretical result, demonstrating a universal scaling of the exponential reduction in the MET (compared to the SSI case) on $\Omega$, for any BSD. The fact that in Fig.~\ref{fig-c}a, $\varepsilon=0.75$, indicates that the bifurcation-limit result may extend beyond its formal region of validity $\varepsilon\ll 1$. Figure~\ref{fig-c}b shows $|\Delta S-\Delta S^b|/\Delta S$ -- the relative error of the accumulated action in the bifurcation limit with respect to the general case -- as a function of $\varepsilon$. As expected, the relative error vanishes as $\varepsilon\to 0$. That the bifurcation limit holds, for various BSDs, as long as $\varepsilon\ll 1$, verifies our theoretical assumption, and further demonstrates the robustness and applicability of the bifurcation result.

We have calculated the mean escape time (MET) from a metastable state in the case where immigration occurs in bursts rather than via arrival of single individuals at a certain rate. This was done using an arbitrary burst size distribution (BSD). We have shown that the effect of bursty immigration can exponentially decrease the MET from a metastable state. Close to the bifurcation limit, when the metastable state and the corresponding unstable state (which serves as the barrier for escape) are close, we have found a simple analytical expression for the decrease in the MET as function of the first two moments of the BSD. Our theory was demonstrated on several prototypical examples of BSDs including the geometric and negative-binomial distributions, appearing in cell biology, as well as scale-free distributions, which describe many social, biological, and communication networks.

We anticipate that step-size noise appearing in problems of similar nature will also strongly influence the long-time behavior of stochastic populations. The formalism presented here can be extended to two additional types of step-size noise: bursty reproduction and death in groups. The former appears in the context of evolutionary biology~\cite{geneticsI,geneticsII}, whereas the latter has been recently observed in nature in populations of \textit{Saiga} antelopes~\cite{saiga}. It would be interesting to study how, \textit{e.g.}, extinction of a population is affected by having a fluctuating offspring number per birth event, or by having an instantaneous removal of an a-priori unknown number of individuals. Here, however, the analysis is expected to be more involved than the case of bursty influx, due to the strong coupling between demographic and step-size noise~\cite{burstyprod}.

Finally, the formalism that we have developed here for static step-size noise makes a first step in the systematic study of step-size noise with arbitrary correlation time. Importantly, the latter may confer new insight as to the role of other types of non-demographic noise, such as extrinsic or environmental noise, see \textit{e.g.} Refs.~\cite{Elowitz,Leigh}, on stochastic populations dynamics.


\bigskip
We thank Baruch Meerson for useful discussions. This work was supported by Grant No. 300/14 of the Israel Science Foundation.



\begin{thebibliography}{99}
\bibitem{cubic} This unlimited growth can be arrested by adding a term proportional to $n^3$ to the death rate leading to a high stable fixed point $n_3>n_2$. For a sufficiently large $n_3$, the MET from $n_1$ to infinity equals in the leading order to the mean switching time between $n_1$ and $n_3$~\cite{ours}.
\bibitem{FnII} Here we have used the mathematical identity
$\sum_{n=0}^{\infty}p^n\sum_{k=0}^nD(k)P_{n-k}= \sum_{k=0}^{\infty}D(k)\sum_{n=k}^{\infty}p^nP_{n-k}$.
\bibitem{fequation} In the scale-free case, we have used the fact that $f(p)\!=\!\sum_{l=1}^\infty \frac{(p\!-\!1)^{l-1}}{l!H_N^{(\gamma-1)}}
\!\sum_{m=1}^l \frac{H_N^{(\gamma-m)}}{m!}\!\left.\frac{d^m}{dy^m}\!\left[\frac{\Gamma(y\!+\!1)}{\Gamma(y\!-\!l\!+\!1)}\right]\right|_{y=0}$,
where $\Gamma[y+1]/\Gamma[y-l+1]=y(y-1)\cdots (y-l+1)$, and $H_N^{(\gamma)}$ is the $N$-th harmonic number of order $\gamma$.
\bibitem{accuracy} Note, that Eq.~(\ref{f}) only accounts for the exponential dependence of $\tau$ on $N$. Subleading-order corrections require a more subtle theoretical treatment in the vicinity of the unstable fixed point~\cite{burstyprod}.
\bibitem{cutoff} To avoid a diverging variance and instant switching, we assume a natural cutoff in the case of scale-free BSD. The maximal burst size is assumed to be $N$ which is on the same order as the distance between the stable and unstable fixed points $n_1$ and $n_2$, for $\alpha={\cal O}(1)$.
\bibitem{FnIX} In Fig.~\ref{fig-b}d we plot $\tau_{SF}$ rather than $\tau_{SF}/\tau_{SSI}$ since the convergence to the SSI case occurs only at $\gamma\to\infty$.
\bibitem{FnVIII} In the case of scale-free BSD we find that in the vicinity of $p_{\mathrm{max}}$, $p-1\sim\varepsilon^2H_N^{(\gamma-1)}/H_N^{(\gamma-2)}$.
\end{thebibliography}
\end{document}